# A Dual Realization of Chua's Chaotic Oscillator Using a Current-Controlled Nonlinear Resistor

Yihang Chen
*Faculty of Electronic Information and Electrical Engineering*
Dalian University of Technology
Dalian, China
yhchendut@gmail.com

Weijie Dong
*Faculty of Electronic Information and Electrical Engineering*
Dalian University of Technology
Dalian, China
dongwj@dlut.edu.cn

Yongping Xie
*Faculty of Electronic Information and Electrical Engineering*
Dalian University of Technology
Dalian, China
xieyp@dlut.edu.cn

*Abstract*—A dual realization of Chua's chaotic oscillator is proposed using current-controlled nonlinear resistors, one linear resistor, one capacitor and two inductors. Two problems are solved. First, unit rescaling is necessary when transforming the standard chaotic equations into circuit equations to ensure that the current units are milliamperes. In addition, the connection and parameters of two current-controlled nonlinear resistors are set to build the required volt-ampere characteristics. The inductor currents show the classical characteristics of being sensitive to the circuit parameters and initial conditions. In addition, experimental verification is performed to demonstrate the feasibility of the circuit. Chua's dual circuit exhibits rich dynamic chaotic features and might bring new applications due to chaotic currents.

## I. Introduction

The chaotic state shows rich dynamic features [1] and is considered to be the fourth state in addition to equilibrium, periodic and quasi-periodic states [2]. Chua's circuit has served as a primary reference in generating and studying bifurcations and chaos [3]. It is widely recognized that Chua's circuit realization of two-op-amp-based voltage-controlled nonlinear resistors has already been accepted as a standard [3]. Although Chua's circuit and similar circuits are simple electronic networks, they exhibit rich bifurcation phenomena and attractors [4]-[5], therefore obtaining wide use in a variety of fields, such as secret communication, automatic control, aerospace, power systems [2][6] and even realms of art [7].

Since Chua's circuit was put forward, diverse methods have been proposed to implement it, such as using voltage-controlled nonlinear resistors [4][8], diodes [9], operational transconductance amplifiers [10], transistors [11] and memristors [12]-[15]. Apart from that, several works have been done to overcome the shortcoming of the inductor by replacing it with other elements [16]-[18]. Recently, circuits that can generate multi-shape chaotic attractors have been proposed [19]-[23], which greatly enrich the topological structures of bifurcations and chaos.

While many studies on voltage-controlled element-based Chua's circuits have been performed, they are far from sufficient for current-controlled element-based circuits. Among the few current-controlled element-based Chua's circuit, almost all are about theoretical analysis [2], [24] or chaotic characteristics simulated on Simulink [25]-[26], and there is no circuit implementation. Considering the necessity of the application of chaotic current output signals in the field [3] and the necessity of being easy to implement[9], we propose an all-new kind of Chua's dual circuit. It has a simple structure, consisting of one capacitor, two inductors, one linear resistor and a pair of series current-controlled nonlinear resistors. Each nonlinear resistor is comprised of one op amp and three linear resistors.

In this paper, the circuit's dynamic equations are derived, and several problems are solved, such as the determination of element parameters and rescaling of the default units. In addition, the effect of initial conditions on the formation of chaos and dual comparisons between the two kinds of nonlinear resistor-based Chua's circuits are discussed. Finally, a rough experimental verification is carried out.

## II. Standard Chua's Chaotic Circuit

This paper only concerns three-order autonomous Chua's chaotic circuits. The general Chua's chaotic equations [1] are given by:

$$\begin{cases} \dfrac{dx}{d\tau} = \alpha\left[y - x - g(x)\right] \\ \dfrac{dy}{d\tau} = x - y + z \\ \dfrac{dz}{d\tau} = -\beta y \end{cases} \quad (1)$$

where $g(x)$ is a five-piecewise function, shown in Fig. 1.

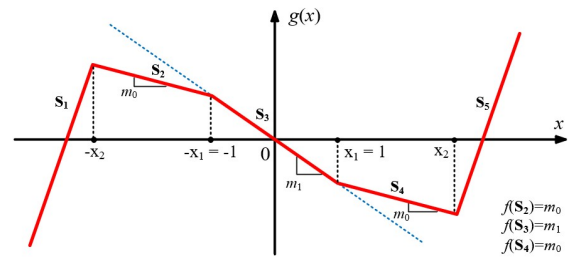

Fig. 1. Typical $g(x) \sim x$ curve, where $f(\mathbf{S_n})$ indicates the slope of segment $\mathbf{S_n}$.

This work was supported by the National Key R&D Program of China under Grant 2019YFB2005900.

This is a revised version of its conference paper version in ICCS 2021, link: https://ieeexplore.ieee.org/document/9697183. Two typing errors are corrected in equation (2) and (5), respectively.

The outermost segments $S_1$ and $S_5$ must lie completely within the first and third quadrants of the $g(x) \sim x$ plane for sufficiently large $|g(x)|$ and $|x|$ [4]. However, segments $S_1$ and $S_5$ do not contribute to the formation of the attractors, and only segments $S_2$ to $S_4$ work to affect the formation of attractors.

Based on this consideration, only segments $S_2$ to $S_4$ are listed in the expressions of $g(x)$, as shown in (2). $m_1$ and $m_0$ in $g(x)$ are assigned as $m_1 = -1.2$ and $m_0 = -0.6$, respectively, in this paper.

$$g(x) = \begin{cases} m_0 x + (m_0 - m_1), & x \leq -1 \\ m_1 x, & |x| < 1 \\ m_0 x - (m_0 - m_1), & x \geq 1 \end{cases} \quad (2)$$

In the classic Chua's chaotic circuit shown in Fig. 2, the $g(x) \sim x$ relation is realized by a voltage-controlled nonlinear resistor $R_x$. The current through $R_x$ varies by the voltage between its two terminals, which is expressed by $i(u_1) \sim u_1$, where $i(u_1) = g(u_1)/R_0$.

### III. DESIGN OF DUAL CHAOTIC

#### A. Current-controlled Nonlinear Resistor

Unlike the voltage-controlled nonlinear resistor, a current-controlled nonlinear resistor means that the voltage between its two terminals is a function of the current that flows through it. A current-controlled nonlinear resistor, which has the volt-ampere characteristics of a three-piecewise linear slope, can be realized by using one op amp and three linear resistors, similar to a voltage-controlled nonlinear resistor. The two types of nonlinear resistors are compared in Fig. 3.

In a voltage-controlled nonlinear resistor, the input voltage is between the noninverting terminal and ground, and the divided output voltage is fed back to the inverting terminal, whereas in a current-controlled nonlinear resistor, the input voltage is connected to the inverting terminal, and the divided output voltage is fed back to the noninverting terminal.

The op amps in both nonlinear resistors work from a negative saturation to a linear state and then to a positive saturation, reciprocating cycle. The three-piecewise $u_R \sim i_R$ relation of a current-controlled nonlinear resistor is:

$$u_R = \begin{cases} R_f i_R + u_{sat}, & i_R \leq -i_p \\ -\dfrac{R_b R_f}{R_a} i_R, & |i_R| < i_p \\ R_f i_R - u_{sat}, & i_R \geq i_p \end{cases} \quad (3)$$

in which $u_{sat}$ is the saturated output voltage of the op amp determined by supply voltage $V_{CC}$ and $u_p$, defined as $\dfrac{R_b}{R_a + R_b} u_{sat}$, is the voltage of inflection points. Attention should be paid that the value of $V_{CC}$ is extremely important. While $V_{CC}$ is changing, $u_{sat}$ will change; therefore, the range of the operating voltage and current will change as well. The volt-ampere characteristic curves of both types of resistors are shown in Fig. 4. The red solid line represents the current-controlled resistor, and the blue dotted line represents the voltage-controlled resistor. There are two different slopes in the characteristics curve, which are $k =$ $R_f$ and $k = -\dfrac{R_b R_f}{R_a}$. The slopes are dependent on the values of linear resistors only.

#### B. Circuit and State Equations

By using duality, an all-new kind of Chua's dual circuit based on current-controlled resistors is proposed, as shown in Fig. 5. Note that two current-controlled nonlinear resistors, each exhibiting three piecewise volt-ampere characteristics, are set in series to form one comprehensive nonlinear resistor (the two terminals of which are node 1 and 2). To make it easier to express and understand, we call it a *combined resistor*. According to Kirchhoff's law, the state equations can be expressed by (4) in SI units as follows:

$$\begin{cases} L_1 \dfrac{di_1}{dt} = R_0(i_2 - i_1) - u(i_1) \\ L_2 \dfrac{di_2}{dt} = R_0(i_1 - i_2) + u_C \\ C \dfrac{du_C}{dt} = -i_2 \end{cases} \quad (4)$$

We define $t = \dfrac{L_2}{R_0}\tau$, $i_1 = x$, $i_2 = y$ and $\dfrac{u_C}{R_0} = z$ and can also obtain (1) after appropriate transformation, in which

$$\alpha = \dfrac{L_2}{L_1}, \beta = \dfrac{L_2}{CR_0^2} \text{ and } g(i_1) = \dfrac{u(i_1)}{R_0}.$$

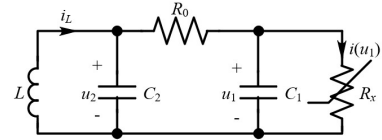

Fig. 2. Chua's circuit based on voltage-controlled resistors.

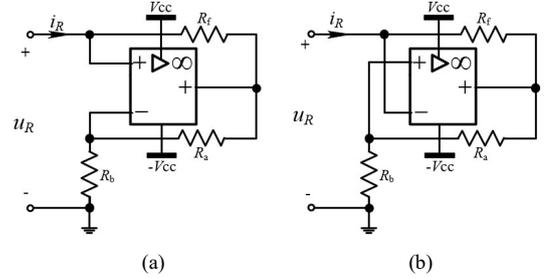

(a)　　　　　　　　　　　(b)

Fig. 3. Circuits of two types of nonlinear resistors: (a) a voltage-controlled nonlinear resistor; (b) a current-controlled nonlinear resistor.

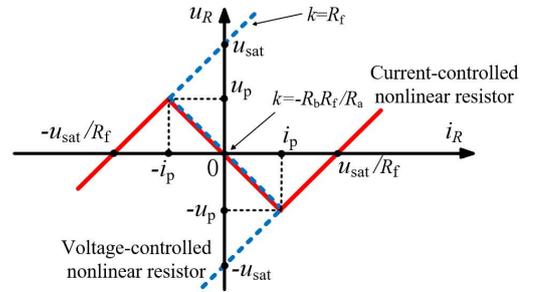

Fig. 4. Comparison of the volt-ampere characteristics of two types of nonlinear resistors.

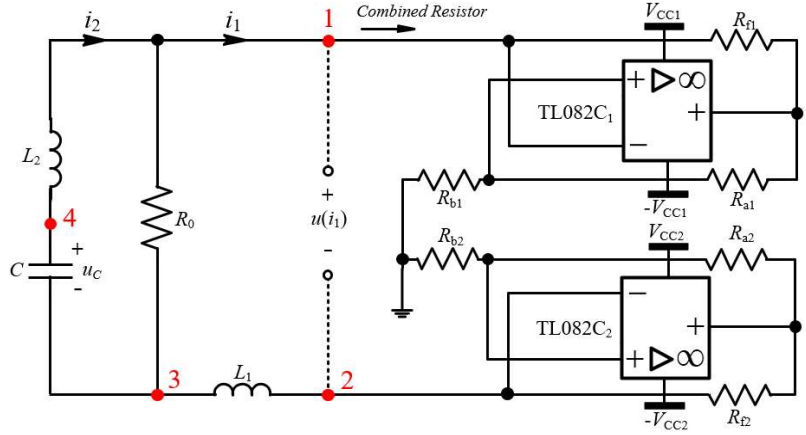

Fig. 5. Chua's dual circuit based on current-controlled resistors.

TABLE I. RESCALING OF UNITS

| Physical quantity | Original unit | Rescaled unit |
|---|---|---|
| Voltage | V | V |
| Current | A | mA |
| Resistance | Ω | kΩ |
| Capacitance | F | mF |
| Inductance | H | kH |

TABLE II. SEGMENTED SLOPES OF $G(I_1)$

| Segment | Expression | Value |
|---|---|---|
| $S_1$ | $(R_{f1} + R_{f2}) / R_0$ | +0.6 |
| $S_2$ | $\left(R_{f2} - \dfrac{R_{b1}R_{f1}}{R_{a1}}\right) / R_0$ | -0.6 ($m_0$) |
| $S_3$ | $-\left(\dfrac{R_{b1}R_{f1}}{R_{a1}} + \dfrac{R_{b2}R_{f2}}{R_{a2}}\right) / R_0$ | -1.2 ($m_1$) |
| $S_4$ | $\left(R_{f2} - \dfrac{R_{b1}R_{f1}}{R_{a1}}\right) / R_0$ | -0.6 ($m_0$) |
| $S_5$ | $(R_{f1} + R_{f2}) / R_0$ | +0.6 |

The functional expression of $g(i_1) \sim i_1$ should be the same as $g(x) \sim x$ in (2). $u(i_1) = R_0 \times g(i_1)$ is the volt-ampere characteristic of the *combined resistor*, which is exhibited as five piecewise segments. Because $u(i_1)$ and $g(i_1)$ are linear mappings, the coordinates of the critical inflection points in $u(i_1)$ can be referred directly from those in $g(i_1)$.

*C. Determination of the Parameters*

Since it is easier and more reasonable to realize the current at the milliampere level than at the ampere level [4], the first step is to assign units of mA to variables $i_1$ and $i_2$ in (4) to transform them to (1). As a result, unit rescaling is required. We keep the voltage unchanged and descale the current by a factor of 1000, from ampere to milliampere. To balance both sides of the equations, the units of other physical quantities should also be rescaled accordingly, as exhibited in TABLE I.

The unit rescaling of resistance helps to determine the parameters of the *combined resistor*. The first step is to determine $g(i_1)$, which is decided by $g(x)$ shown in (2). Based on $g(i_1)$, $u(i_1)$ can be deduced. By using linear superposition, the slopes of different segments of $g(i_1)$ can be given by linear resistors, which are shown in TABLE II. $R_0$ is the denominator in the slopes; therefore, slopes have no units. As shown in Fig. 1, the piecewise slopes are $m_1$=-1.2 and $m_0$=-0.6 and the inflection point is $x_1$=1 (mA). Given $R_0$ = 5 kΩ, the volt-ampere relation of the *combined resistor* is given by (5):

$$u(i_1) = R_0 \times g(i_1) = 5 \times g(i_1) = 5 \times \begin{cases} -0.6i_1 + 0.6, & i_1 \leq -1\text{mA} \\ -1.2i_1, & |i_1| < 1\text{mA} \\ -0.6i_1 - 0.6, & i_1 \geq 1\text{mA} \end{cases} \quad (5)$$

Consequently, the parameters of the linear resistors in Fig. 5 can be calculated as:

$R_{f1}$ = 1.5 kΩ, $R_{a1}$ = 6 kΩ, $R_{b1}$ = 18 kΩ, $V_{CC1}$ = 18 V;
$R_{f2}$ = 1.5 kΩ, $R_{a2}$ = 10 kΩ, $R_{b2}$ = 10 kΩ, $V_{CC2}$ = 4.6 V;

In (5), $R_0$ is measured in kΩ, and $g(i_1)$ and $i_1$ are measured in mA; therefore, $u(i_1)$ can finally be determined in volts, as shown in Table I. Other parameters in Fig. 5 are assigned as $L_1$ = 5× $10^{-5}$ kH, $L_2$ = 4×$10^{-4}$ kH and $C$ = 1.28×$10^{-6}$ mF, to maintain $\alpha$ = 8 and $\beta$ = 12.5. The units kΩ, kH and mF are the results of unit rescaling to ensure that the unit of current is mA. A pair of TL082Cs are used as the op amps in this circuit, as shown in Fig. 5. The following points should be given extra attention when building the *combined resistor*.

First, the ground node should be set between the two op amps that are set in series to ensure that both op amps can be connected to the ground and obtain the correct feedback voltage.

Second, to form $g(i_1)$, as shown in Fig. 1, the voltage of each nonlinear resistor should be added together and then divided by $R_0$. Hence, in segments $S_2$, $S_3$ to $S_4$ in Fig. 1, TL082C$_1$ should only work at its linear state and therefore be supplied with a higher voltage $V_{CC1}$ = 18 V, whereas TL082C$_2$ supplied with $V_{CC2}$ = 4.6 V works at all its linear and saturation states. In detail, in segments $S_1$ and $S_5$, both op amps work in the saturation regions. In segments $S_2$ and $S_4$, TL082C$_1$ works in the linear region, and TL082C$_2$ works in saturation regions. In segment $S_3$, both op amps work in the linear region.

Third, the element parameters of each linear resistor are of great importance. The value of $R_f$ is positive, but $R_b R_F/R_a$ has a negative changing trend towards the volt-ampere characteristics of the nonlinear resistors. Additionally, as the linear resistors change, the inflection points and intercepts of equations $g(i)$ change as well. This would affect the formation of chaotic attractors.

### D. Bifurcations and Chaos

Using Multisim 14, we investigated the chaotic characteristics of the proposed circuit shown in Fig. 5. Two 1 Ω sampling resistors are in series with inductors to convert $i_1$ and $i_2$ into voltage signals to enable the oscilloscope to detect the current signals. $\alpha$ and $\beta$ are not the only factors that impact the dynamic performance but also $R_0$. Keeping other parameters unchanged, we change $R_0$ and obtain the results as follows. We obtain the double-scroll attractor phase space trajectory of $i_1 \sim i_2$ when $4626\,\Omega \leq R_0 \leq 5970\,\Omega$ and the initial values of $i_1$, $i_2$, and $u_C$ are set to zero, as shown in Fig. 6. We set $4526\,\Omega \leq R_0 \leq 4625\,\Omega$, and furthermore get a single-scroll attractor phase space trajectory when the initial conditions remain zero, as shown in Fig. 7. As $R_0$ decreases, the phase space trajectory changes from a double-scroll attractor to a single-scroll attractor and finally shows no chaos. Therefore, it is critical to control the value of $R_0$ as well if special attractors are wanted.

### E. Effect of the Initial Conditions

In the simulation by Multisim 14, chaotic characteristics are also found to be sensitive to the initial values [1]. The circuit cannot generate chaos when the initial conditions are not in a suitable range.

We mainly observe the effect of the initial voltage of the capacitor. Therefore, we set the circuit parameters so that they are the same as in section C, and the initial currents of the two inductors are set to zero. When $-11.91\,\text{V} \leqslant u_C(0) \leqslant 11.89\,\text{V}$, the circuit can generate chaotic attractors, which shows that our circuit has a broad tolerance of initial voltage. When $u_C(0)$ is around the corner of the boundary values, circular orbits would first form, and then chaotic attractors would come into being, as shown in Fig. 8.

### F. Parameter Reselection for Experimental Validation

The current-controlled nonlinear resistor-based Chua's circuit is verified by simulations as shown above, on the basis of which a hardware circuit is obtained. Due to the difficulty of lab implementation of large capacitance and inductance, it is necessary to reduce the values of $L_1$, $L_2$ and $C$.

We maintain $R_0 = 5$ kΩ and initial conditions of zero and scale down the values of the capacitor and the inductors under the condition $\alpha = 8$, and $\beta$ slowly reduces from 12.5. To build the experimental circuit, the values of the elements are selected as shown below, referring to Fig. 5. The op amps are selected as TL082CDs.

$$C = 86 \text{ pF, parallel of 82 pF, 2 pF and 2 pF};$$
$$L_1 = 1.2 \text{ mH}, L_2 = 10 \text{ mH}, R_0 = 5 \text{ kΩ};$$
$$R_{f1} = 1.5 \text{ kΩ}, R_{a1} = 6 \text{ kΩ}, R_{b1} = 18 \text{ kΩ};$$
$$R_{f2} = 1.5 \text{ kΩ}, R_{a2} = 10 \text{ kΩ}, R_{b2} = 10 \text{ kΩ};$$

All the elements of our experimental circuit except the sources and the oscilloscope are shown in Fig. 9. Red wires refer to positive power supplies, blue wires refer to negative power supplies and black wires refer to the ground. Note that the two TL082CDs should be supplied with two different volts.

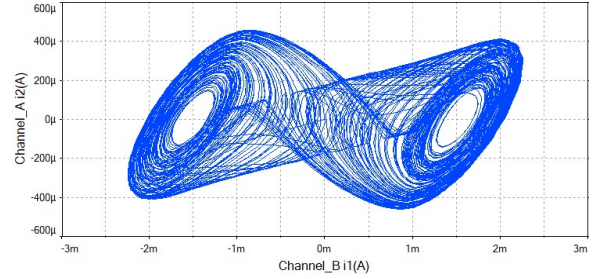

Fig. 6. A double-scroll attractor of $i_2 \sim i_1$ phase space trajectory of current-controlled-resistor-based Chua's circuit when $R_0$=5000Ω.

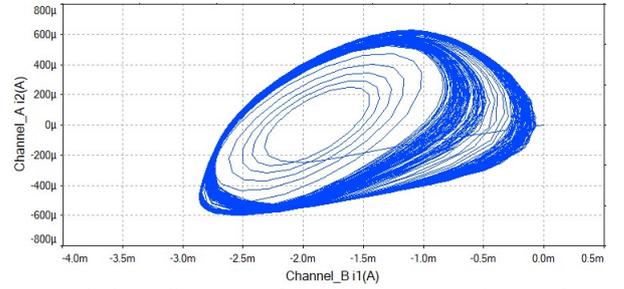

Fig. 7. A single-scroll attractor of $i_2 \sim i_1$ phase space trajectory of current-controlled-resistor-based Chua's circuit when $R_0$=4600Ω.

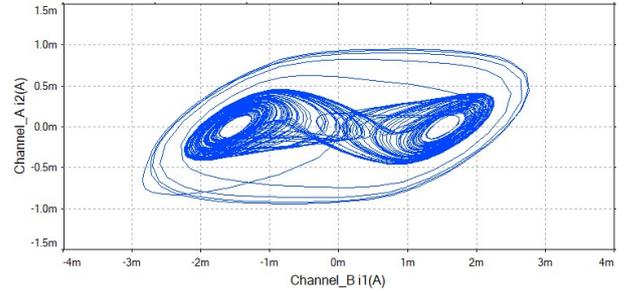

Fig. 8. A double-scroll attractor of $i_2 \sim i_1$ phase space trajectory when $R_0$=5000Ω and $u_C(0)$ is -11.91 V. The trajectory would move from the outside circular orbits into the inner chaotic double-scroll attractor.

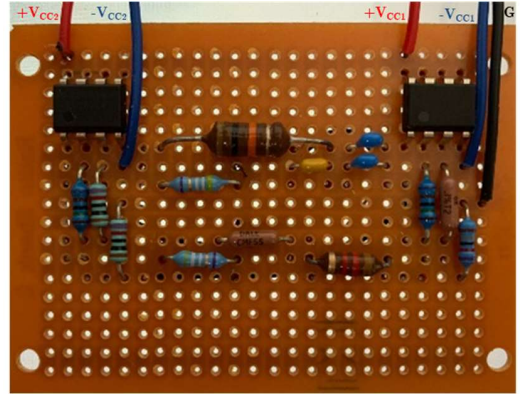

Fig. 9. Picture of the experimental circuit board.

## IV. EXPERIMENTAL RESULTS

If directly measure the inductor currents, then oscilloscope with current probes are needed. If use resistors to sample the currents, then isolated oscilloscope are needed. Alternatively, a pair of node voltages $u_3$ and $u_4$, as indicated in Fig. 5, are measured, where $u_3$ is the voltage of node 3 and $u_4$ is the voltage of node 4.

Given $V_{CC1}$ = 18 V and $V_{CC2}$ = 5 V, the simulated phase diagram of $u_4 \sim u_3$ is shown in Fig. 10(a), which shows as a double-scroll attractor. The simulated time-domain curves are shown in Fig. 10(b). Then with the same power settings, the experimental results and the experimental photos are exhibited in Fig. 10(d), Fig. 10(e) and Fig. 10(c). Here, a DH1723-1 Power Supply and a Tektronix DPO 2012 Oscilloscope are used. Similarly, simulation and experimental results of node voltage $u_2$ and $u_4$ are shown in Fig. 11.

The experimental results, which are in good agreement with the theoretical prediction and circuit simulation, verify the feasibility of the proposed chaotic circuit.

## V. CONCLUSION

An all-new implementation of Chua's dual circuit based on current-controlled nonlinear resistors is proposed. Simulation on Multisim 14 and laboratory verification are performed to demonstrate the feasibility of the circuit. Several characteristics of Chua's dual circuit and its comparison with the classic circuit are shown. The circuit has two main features. First, the circuit is realized based on a current-controlled nonlinear resistor, that is, the *combined resistor* in Fig. 5. Second, the circuit has few components and a simple structure so that we can implement it in the laboratory. The following details still require attention:

1) The difference in the connection of noninverting and inverting terminals between voltage-controlled and current-controlled nonlinear resistors is remarkable.

2) Expectation values of the elements should be commercially available.

3) Choices of initial values, which affect the formations of chaotic attractors.

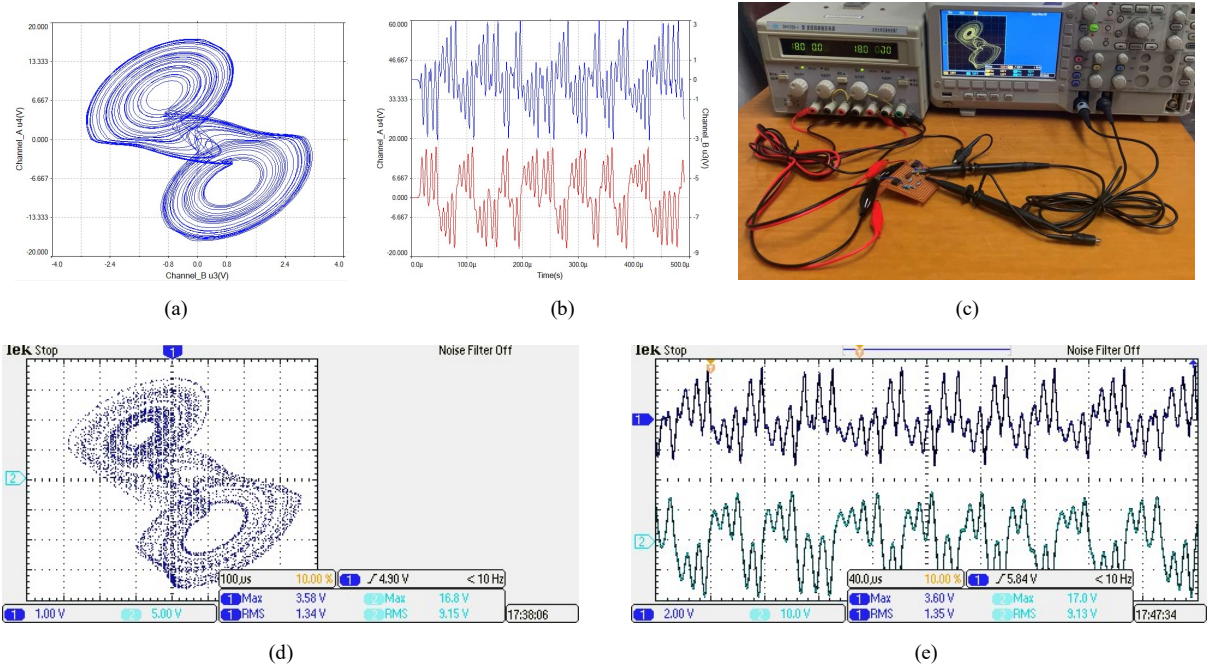

Fig. 10. Chaotic curves of $u_3$ and $u_4$ of current-controlled nonlinear resistor-based Chua's circuits: (a) simulated double-scroll attractor; (b) simulated time-domain curves (chA(lower): $u_4$, chB(upper): $u_3$); (c) experimental photo; (d) measured double-scroll attractor; (e) measured time-domain curves (ch1: $u_3$, ch2: $u_4$).

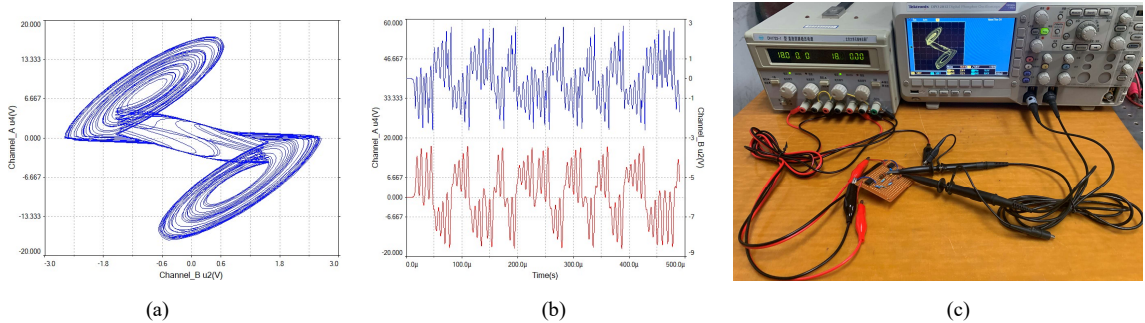

(a)    (b)    (c)

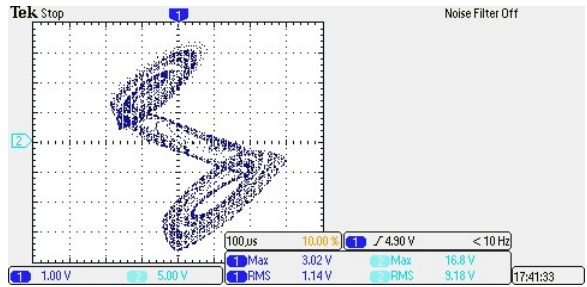
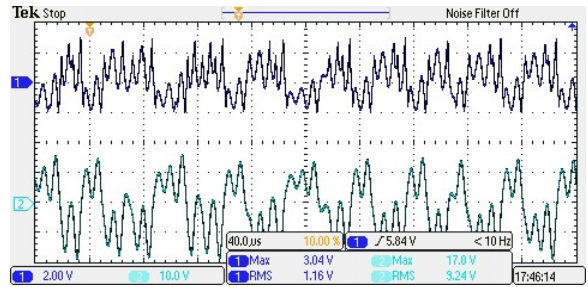

(d)               (e)

Fig. 11. Chaotic curves of $u_2$ and $u_4$ of current-controlled nonlinear resistor-based Chua's circuits: (a) simulated double-scroll attractor; (b) simulated time-domain curves (chA(lower): $u_4$, chB(upper): $u_2$); (c) experimental photo; (d) measured double-scroll attractor; (e) measured time-domain curves (ch1: $u_2$, ch2: $u_4$).